\documentclass[preprint,12pt,num,sort&compress]{elsarticle}

\usepackage{amssymb,amsmath}
\usepackage{xcolor,hyperref}
\usepackage[bottom]{footmisc}
\usepackage{slashed}
\usepackage{wrapfig}
\definecolor{weinbergblue}{RGB}{10, 10, 100}
\definecolor{darkblue}{rgb}{0.0, 0.0, 0.55}
\definecolor{glasgowblue}{RGB}{0, 45, 76}

\journal{Physics Letters B}
\hypersetup{colorlinks=true, citecolor=magenta, urlcolor=magenta, linkcolor=magenta}
\begin{document}
\flushbottom
\allowdisplaybreaks

\begin{frontmatter}

\title{A $B-$anomaly motivated $Z^\prime$ boson at the energy and precision frontiers} 

\author[1]{Ben Allanach}
\ead{ben.allanach.work@gmail.com}
\affiliation[1]{organization={DAMTP, University of Cambridge},
  addressline={Centre for Mathematical Sciences, Wilberforce Road},
  postcode={CB3 0WA},
  city={Cambridge},
  country={United Kingdom}}
\author[2]{Christoph Englert}
\ead{christoph.englert@manchester.ac.uk}
\affiliation[2]{organization={Department of Physics and Astronomy, University of Manchester},
  addressline={Schuster Building, Oxford Road},
  postcode={M13 9PL},
  city={Manchester},
  country={United Kingdom}}
\author[3]{Wrishik Naskar}
\ead{wrishik.naskar@glasgow.ac.uk}
\affiliation[3]{organization={School of Physics and Astronomy, University of Glasgow},
  addressline={Kelvin Building, University Avenue},
  postcode={G12 8QQ},
  city={Glasgow},
  country={United Kingdom}}

\begin{abstract}
TeV-scale $Z^\prime$ bosons with family-dependent couplings can explain some anomalies inferred from $B-$meson measurements of processes involving the $b \rightarrow s \ell^+\ell^-$ transition. A $Z^\prime$ originating from kinetically-mixed spontaneously broken $U(1)_{B_3-L_2}$ gauge symmetry has been shown to greatly ameliorate global fits~\cite{Allanach:2024ozu} in a `flavour-preferred' region of parameter space. We provide an exploration of this region at the high luminosity (HL-)LHC with particular attention to which signals could be verified across different discovery modes. Even if the HL-LHC does not discover the $Z^\prime$ boson in a resonant di-lepton
channel, a FCC-ee $Z$-pole run would detect oblique corrections to the electroweak precision observables (EWPOs). Changes due to $Z^\prime$-induced non-oblique corrections are unlikely to be detected, to within experimental precision. In any case, the extended discovery potential offered by a 100 TeV $pp-$collider would afford sensitivity to the entire flavour-preferred region and enable a fine-grained and forensic analysis of the~model.
\end{abstract}

\end{frontmatter}

\renewcommand{\thefootnote}{\fnsymbol{footnote}}
\section{Introduction}
\label{sec:introduction}
There exists an interplay between
two discovery-led research pillars of
the current high-energy particle physics programme: precision flavour
investigations which require experimental and theoretical control
and searches for beyond the Standard Model (BSM) states at the highest
attainable energies, notably at the Large Hadron Collider (LHC). 
Searches for new particles in `bump-hunt' analyses do not typically require
dedicated experimental and theoretical control; they are data-driven discovery
methods.
The hypothesis of a certain model BSM can link precision flavour
investigations to the search programme (see e.g.~\cite{Bordone:2021cca,Bissmann:2019gfc,Grunwald:2023nli,Englert:2024nlj,Davighi:2024syj} for similar studies). 
This not only enables us to make concrete discovery predictions based on observed flavour anomalies~\cite{Capdevila:2023yhq} but also suggests a multi-faceted exploration programme characterised by checks and balances.

We demonstrate the phenomenological power of such a programme by investigating
a concrete model that changes
Standard Model predictions for various measurements involving flavour-changing $B-$meson decays, the
malaphoric $U(1)_{B_3-L_2}$ model~\cite{Allanach:2024ozu}.
In this model, the TeV-scale $Z^\prime$ has interactions with $e^+e^-$, with
$\mu^+ \mu^-$ and with $\bar b s$ and $\bar s b$. It therefore
introduces a BSM contribution which interferes with the SM $b \rightarrow s
\ell^+ \ell^-$ amplitude\footnote{We denote $\ell$ to be a charged lepton in the usual
experimental parlance, i.e.\ either muons or electrons.}. 
Ref.~\cite{Allanach:2024ozu} shows that the model ameliorates a SM fit to flavour
observables: in a global fit to hundreds of observables, an improvement of 38.2
units of total $\chi^2$ for three effective fit parameters was found.
LEP2 di-lepton production constraints on the differential cross-sections,
EWPOs and lepton flavour universality
constraints are included and are simultaneously well fit.

The lowest hanging fruit of the malaphoric $U(1)_{B_3-L_2}$ model as far as
searches are concerned is the bump-hunt search in di-lepton
channels~\cite{Allanach:2024nsa}; depending upon parameters, this could
discover the predicted $Z^\prime$ boson.
Here, we build on the search to provide
a more comprehensive 
analysis of the discovery potential at the high-luminosity (HL-)LHC and
clarify how combinations of channels with different statistical sensitivity
can aid the process of determining the model parameters.
In the event that the HL-LHC does not report a discovery of direct $Z^\prime$
production, it will provide a lower limit on the mass of the $Z^\prime$
boson $M_{Z^\prime}$ as a function of
$g_X$, the $U(1)_{B_3-L_2}$ gauge coupling. An emergent question is then
whether a potential future precision $e^+e^-$ collider could provide additional
sensitivity beyond the direct exclusion constraints established at the LHC. By
relying on the underlying model BSM, we can address this question by investigating BSM electroweak corrections, in particular of a precision $Z$-pole programme.

We organise this note as follows. In Sec.~\ref{sec:model}, we briefly sketch
salient features of the malaphoric $U(1)_{B_3-L_2}$ model. Specific care is
also given to a detailed discussion of rare $Z^\prime$ final states, to inform
the discovery projection of the HL-LHC. This is discussed in
Sec.~\ref{sec:hl-lhc} where we compare the flavour-preferred parameter region
with the dominant discovery modes of the model, extending the results of
Ref.~\cite{Allanach:2024ozu}. Building on the HL-LHC
sensitivity prospects for the considered scenario, we turn to precision
$e^+e^-$ measurements in Sec.~\ref{sec:fcc-ee}. Concretely, we investigate
helicity asymmetries at one-loop accuracy to navigate the sensitivity of a $Z$
pole programme to the
new physics of the malaphoric $U(1)_{B_3-L_2}$ model. We close with conclusions in Sec.~\ref{sec:conc}.

\section{The Model}
\label{sec:model}
\noindent In the malaphoric $B_3-L_2$ model,
the SM is extended by a $U(1)_{B_3-L_2}$ gauge boson, three right-handed
neutrino fields and a SM-singlet complex scalar $\theta$ with unit $U(1)_{B_3-L_2}$
charge whose vacuum expectation value $v_X \sim
{\mathcal O}(\text{TeV})$ spontaneously breaks $U(1)_{B_3-L_2}$. The $U(1)_{B_3-L_2}$ charges
of all
fields are set to be proportional to\footnote{The constant of
    proportionality is $3$ in order to make all $X_\psi$ integers.} third family baryon number minus second family
number. The
$Z^\prime$ couples to all SM di-fermion pairs via 
the kinetic mixing parameter $\epsilon$ in the Lagrangian
\begin{multline}
  {\mathcal L}_{XB} = -\frac{1}{4} X_{\mu\nu} X^{\mu \nu} +
  g_X^2 X_H^2 |H^\dag H|  X_\mu X^\mu +
  \frac{1}{2} M_X^2
  X_\mu   X^\mu\\
  - \frac{\epsilon}{2} B_{\mu \nu}X^{\mu \nu}- X_\mu J^\mu - B_\mu j^\mu,
  \label{eq:LXB}
\end{multline}
where $X_{\mu \nu}$ is the $B_3-L_2$ field strength
tensor and $B_{\mu \nu}$ is the hypercharge field strength tensor.
$J^\mu=g_X \sum_{\psi^\prime} \bar{\psi}^\prime X_{\psi^\prime} \gamma^\mu \psi^\prime$ is the
$B_3-L_2$ current, where the sum is over the chiral Weyl fermions in the
interaction basis $\psi^\prime$ and $X_{\psi^\prime}$ is the $B_3-L_2$ charge of
$\psi^\prime$ (normalised such that the charge of a third-family quarks is $+1$). $j_\mu$ is the hypercharge current, as defined in Ref.~\cite{Allanach:2024ozu}.
After rotation to the mass eigenbasis of the neutral gauge boson fields ${\bf
P}_\mu:=(A_\mu,\ Z_\mu,\ Z^\prime_\mu)^T$, one obtains the following important
Lagrangian terms
\begin{equation}
  {\mathcal L}_{imp} = -\frac{1}{4} {\bf P}^T_{\mu \nu} {\bf P}^{\mu\nu}
  +\frac{1}{2} {\bf P}_\mu M^2_{\bf P} {\bf P}^\mu
  - \sum_{\psi^\prime} \bar \psi^\prime\, {\bf \hat l}^T C \slashed{\bf P} \psi^\prime,
  \end{equation}
where $C$ is the 3 by 3 non-unitary matrix given in~\ref{app:c}, $M_{\bf P}^2=\text{diag}(0, M_Z^2,$ $M_{Z^\prime}^2)$, and $M_Z$ and
$M_{Z^\prime}$ are the physical $Z^0$ and $Z^\prime$ boson masses,
respectively. Working in the approximation~\cite{Allanach:2024nsa}
  $M_Z/M_{Z^\prime}\ll 1$, 
\begin{equation}
M_{Z^\prime}
= \frac{\sqrt{1+ s_w^2 \epsilon^2}}{\sqrt{1-\epsilon^2}} M_X,
\end{equation}
where $s_w=\sin \theta_w$, the sine of the weak mixing angle and we have
introduced the three-vector 
${\bf \hat l}:= \left(g' Y_{\psi^\prime},\  g \hat L,\ g_X 
X_{\psi^\prime}\right)^T \label{l} $, where $g'$ is the hypercharge gauge coupling, $g$ the
$SU(2)$ gauge coupling, $Y_{\psi^\prime}$ is the hypercharge of $\psi^\prime$ and $\hat L$ is an
operator that annihilates singlets of $SU(2)_L$ but
returns an eigenvalue of $T_3/2$ when acting on left-handed doublet fields. $T_3$ is
the third Pauli matrix.

The $Z^\prime$ has
enhanced couplings to di-muons and also to third family quark pairs. Mismatches between
the gauge
interaction basis and the mass basis of the left-handed strange and
left-handed bottom fields mean that an interaction term proportional to 
$g_X \sin \theta_{sb}\bar b \slashed{Z}^\prime P_L s + H.c.$ is present in the
Lagrangian density, where $P_L$
is the left-handed projection matrix for 4-component Dirac
fermions. $\theta_{sb}$ parameterises the mismatch between the gauge and mass
eigenstates of the $s_L$ and $b_L$ fields. 
With these
couplings in place, the $Z^\prime$ mediates an effective interaction
$(\bar b \gamma_\mu P_L s)(\mu^+ \mu^-)$ as well as $(\bar b \gamma_\mu P_L s)(e^+e^-)$. New physics contributions to these two operators are preferred~\cite{Alguero:2023jeh,Allanach:2023uxz} by
some state-of-the-art estimates of the $B-$meson decay channel's
amplitude~\cite{Gubernari:2022hxn,Parrott:2022zte}, although there is debate
about whether all SM effects have been adequately accounted
for~\cite{Ciuchini:2022wbq}.
It was found in Ref.~\cite{Gubernari:2022hxn} that the
best global fit point is $\hat \epsilon = -0.92$, $\hat g_X=0.078$ and
$\theta_{sb}=-0.17$, where
\begin{equation}
  \hat \epsilon:=\epsilon \text{(3 TeV)}/ M_X, \qquad \hat g_X:=g_X \text{(3
  TeV)}/M_X \label{pars}
\end{equation}
$\hat g_X$ is in the domain $0.03-0.23$ in the 95$\%$ confidence level (CL) good-fit region, whereas
$\hat \epsilon$ is less than $-0.1$~\cite{Allanach:2024ozu}.
Direct $Z^\prime$ production signatures at colliders are insensitive to the
value of $\theta_{sb}$ to a good approximation~\cite{Allanach:2024nsa}, and we
set it to zero in our studies throughout the rest of this paper. 

\begin{figure}[!t]
  \centering
  \parbox{0.54\textwidth}{\includegraphics[width = 0.54\textwidth]{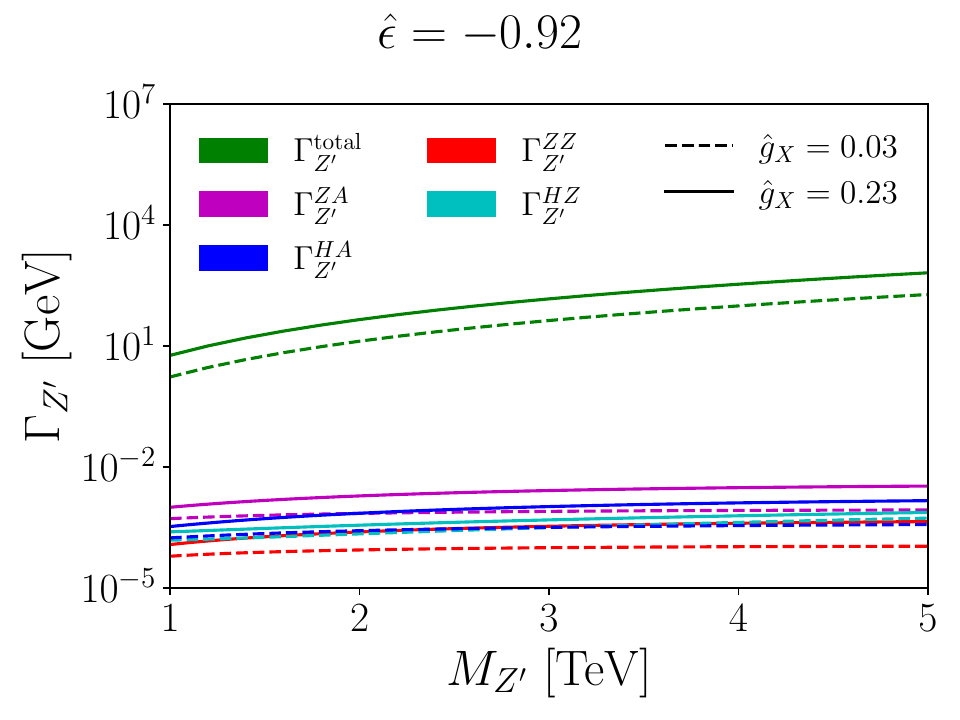}}
 \hfill
  \parbox{0.4\textwidth}{\vspace{3.3cm}\caption{\label{fig:loop_decays} Representative loop-induced decays compared to the unsupressed decay modes exploited further in this work.}}
\end{figure}
%
\subsection*{Comment on loop-induced production and decays}
The one-loop mediated decays $Z^\prime\to ZZ, Z\gamma, Zh$, etc., could
provide an additional \emph{a priori} opportunity to observe the $Z^\prime$ at hadron
machines. The Furry~\cite{Furry:1937zz} and Yang~\cite{Yang:1950rg}
  theorems, and their extensions to decays into massive final states~\cite{Keung:2008ve}, highly constrain the decay amplitudes (see also Refs.~\cite{Michaels:2020fzj,Davighi:2021oel}). They directly imply vanishing $Z^\prime\to \gamma\gamma,gg$ decays. 
Non-vanishing amplitudes are controlled by the $Z^\prime$ axial vector couplings and
the small mass splitting of the $SU(2)_L$ fermion doublet, which leads to an overall suppression of the loop-induced decays.
To understand this more quantitiatively, 
we have performed a full computation of representative loop-induced
leading order decays and compare them to tree-level $Z^\prime$ decay
phenomenology in the parameter region selected by flavour observations in
Fig.~\ref{fig:loop_decays}. As is visible there, the loop-induced decays are
typically two to three orders of magnitude below the dominant $Z^\prime$
decays. This also means that production mechanisms via gluon fusion (necessarily in
association with additional jets to avoid symmetry selection rules) are highly suppressed. 
\section{Phenomenological opportunities at the HL-LHC and high precision lepton facilities}
\label{sec:pheno}
\subsection{$Z^\prime$ at the HL-LHC and FCC-hh: muon signals and multi-jet cross checks}
\label{sec:hl-lhc}
\begin{figure}[!b]
  \centering
  \includegraphics[width = 0.99\linewidth]{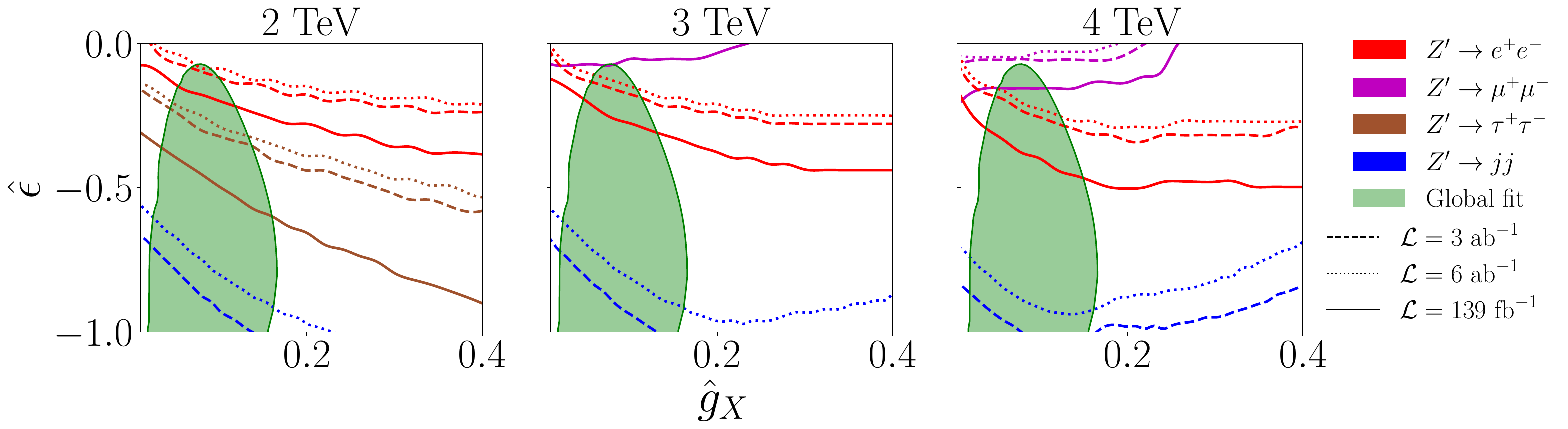}
  \caption{Bump-hunt bounds on the flavour-motivated malaphoric $B_3-L_2$
    model parameter space for $M_{Z^\prime} = 2,3,4~\text{TeV}$. The region
    below each contour is sensitive to the 95$\%$~CL.
   139 fb$^{-1}$ curves are LHC bounds whereas the 3 and 6 ab$^{-1}$
   curves are the projected sensitivities at HL-LHC. For $M_{Z^\prime}=2~\mathrm{TeV}$, the $Z^\prime \rightarrow \tau^+\tau^-$ constraints are shown as an additional channel; higher-mass $\tau^+\tau^-$ limits are not included due to the absence of cross-section upper bounds.
   $Z^\prime \rightarrow \mu^+\mu^-$ curves in the left panel are in the $\hat
   \epsilon > 0$ region.
   The region of 95$\%$CL resulting from the global fit in Ref.~\cite{Allanach:2024ozu} is
     shown as the coloured region. 
   \label{fig:collider} }
\end{figure}

To obtain collider constraints on the model, we calculate the constraints from dedicated
`bump-hunt' searches across multiple final states. By a similar methodology, we acquire sensitivity estimates for future
collider runs. Our analysis extends the work
of Ref.~\cite{Allanach:2024nsa}, which focused on di-electron and di-muon
final states, by incorporating di-jet, and di-tau resonance searches as well.
The model is
implemented in \texttt{FeynRules}~\cite{Christensen:2008py,Alloul:2013bka}, which is subsequently
interfaced with \texttt{MadGraph5\_aMC@NLO}~\cite{Alwall:2014hca} via a
\texttt{UFO}~\cite{Degrande:2011ua} model taken from
Ref.~\cite{Allanach:2024nsa}. We generate events for the production of the
$Z^\prime$ via quark-antiquark annihilation (dominantly $b \bar b$), with
subsequent decays
into di-jet (including $b \bar{b}$ final states), di-muon, and di-electron final states, at a centre-of-mass energy
$\sqrt{s} = 13~\text{TeV}$. In addition, we also include the $\tau^+\tau^-$ final state for the benchmark $M_{Z^\prime} = 2~\mathrm{TeV}$\footnote{Refs.~\cite{ATLAS:2020zms,ATLAS:2025oiy} do not provide cross section upper limits for $\tau^+\tau^-$ resonances at $M_{Z^\prime} = 3,4~\mathrm{TeV}$, preventing a direct recast for these masses.}, which allows us to cross-check the model's sensitivity in the third-generation lepton sector.
We scan over a domain of values for
($\hat{g}_X$,$\hat{\epsilon}$) consistent with constraints from the previous
literature~\cite{Allanach:2024nsa,Allanach:2024ozu}\footnote{The HL-LHC
    will increase the precision on the measurements of $b \rightarrow s \ell^+
    \ell^-$ quantities. For example, the uncertainties in the measurement of
    $P_5^\prime$ in the bin of di-muon mass squared between 4 and 6 GeV$^2$,
    is 
      expected to decrease by a factor of about 7 if 300~fb$^{-1}$ of
      integrated luminosity is
      recorded by LHCb~\cite{LHCTalk}. However, the uncertainties on new physics
      parameter space are not expected to decrease by the same factor because
      they are dominated by theoretical uncertainties in the
      $B$-meson decay predictions.}, for three benchmark mass
values: $M_{Z^\prime} = 2, 3, 4$ TeV.
The 95$\%$ upper limits on the LHC
cross-section times the branching ratios for di-jet, di-muon, di-electron, and di-tau
final states for the respective benchmark points are taken from recent LHC
analyses~\cite{Maguire:2017ypu,ATLAS:2019erb,CMS:2019gwf,ATLAS:2020zms,ATLAS:2025oiy} that provide
model-independent constraints on potential new physics contributions at an
integrated luminosity\footnote{Our results are consistent with those of Ref.~\cite{Allanach:2024nsa}, which found
  $M_{Z^\prime}>2.8$ TeV from current LHC di-muon searches, with sensitivity up to
  $M_{Z^\prime}=4.2$ TeV at the HL-LHC.} $\mathcal{L} = 139~\text{fb}^{-1}$.
To extrapolate these
limits to estimate the High-Luminosity LHC (HL-LHC) sensitivity, we apply a $\sqrt{\mathcal{L}}$
scaling to the expected upper bound on the 95$\%$ CL production
cross-section, which allows us to estimate the sensitivity of future HL-LHC
runs at $\mathcal{L} = 3,6~\text{ab}^{-1}$ to the malaphoric $U(1)_{B_3-L_2}$ model. The collider
sensitivities in the $(\hat{g}_X,\hat{\epsilon})$ parameter space for the
different decay channels for the chosen mass benchmarks are shown on
Fig.~\ref{fig:collider}.
The region of good-fit coming from a global fit to of measurements of
  $e^+e^-\rightarrow \ell^+\ell^-$
  differential cross-sections at LEP2, to EWPOs and to $B-$meson 
  data is shown. 
%
%
\begin{figure}[!t]
  \centering
\parbox{0.55\textwidth}{\includegraphics[width = 0.54\textwidth]{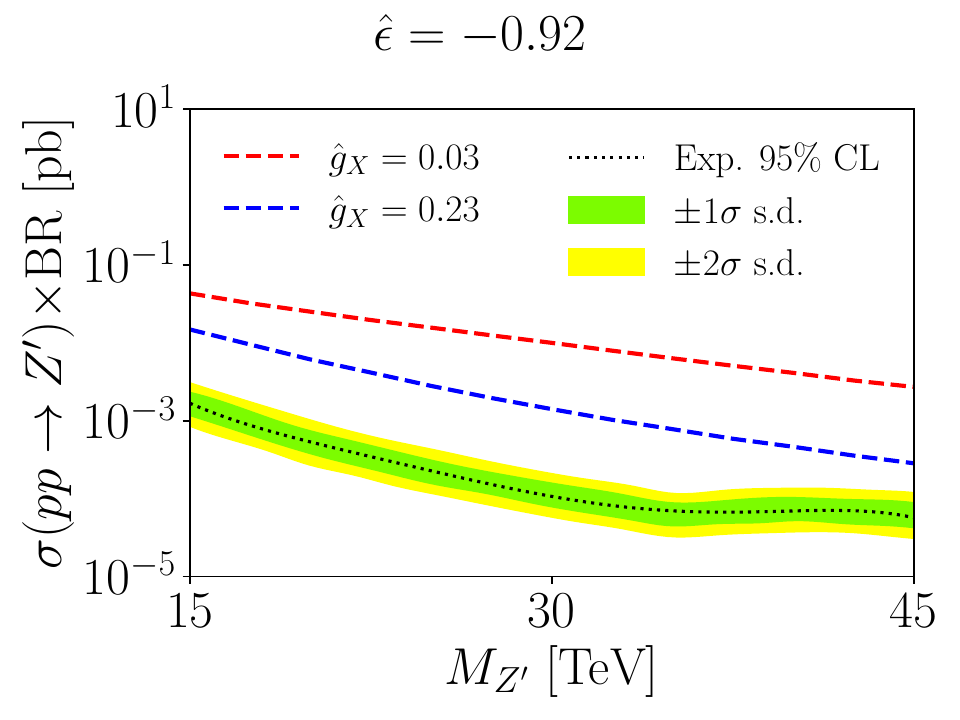}}
\hfill
\parbox{0.4\textwidth}{\vspace{2.05cm}\caption{95\% CL sensitivity for a
    di-jet invariant mass bump hunt at the FCC-hh for 100 TeV collisions at an integrated
    luminosity $\mathcal{L} = 30~\text{ab}^{-1}$. The entire 95$\%$~CL
    global-fit parameter space can be probed. \label{fig:fcc}}} 
\end{figure}
The di-jet resonance searches provide the weakest exclusions and
sensitivity of the channels we study and are not sensitive to much of the
interesting parameter space. 
For $M_{Z^\prime}=2~\mathrm{TeV}$, the $\tau^+\tau^-$ channel yields
  stronger exclusions than the di-jet searches, although still weaker than the
  di-electron limits, as expected from the experimental challenges associated
  with $\tau$ reconstruction. While the experimental analyses considered do
  not provide upper limits on cross sections for $M_{Z^\prime}=3,4~\mathrm{TeV}$, we expect that the $\tau^+\tau^-$ constraints would continue to lie above the corresponding di-jet bounds, but below the di-electron limits, following the trend observed at 2~TeV.

We extend our analysis to the FCC-hh ($\sqrt{s} =
100~\text{TeV}$) to evaluate its reach and to see if the di-jet channel will
be of more use in such a context. We examine the FCC-hh sensitivity to
di-jet resonance searches for the best-fit flavor constraints ($\hat{\epsilon}
= -0.92$, $\hat{g}_X = (0.03,0.23)$). Setting the best-fit flavour constraints,
we scan over masses and generate cross-sections for $M_{Z^\prime} =
20-40~\text{TeV}$. The resulting cross-sections, along with the $95\%$
C.L. constraints on the projected upper bounds on di-jet resonance
cross-sections at the FCC-hh~\cite{Helsens:2019bfw}, are shown in
Figure~\ref{fig:fcc}. These demonstrate that the FCC-hh should easily be sensitive
to all of the 95$\%$CL preferred parameter region\footnote{We note that
    at such high values of $M_{Z^\prime}$ (and therefore high values of $g_X$
    in the good-fit region), 
    one could also check bounds from LHC Drell-Yan cross-section measurements at
    high masses, such as was done in Ref.~\cite{Greljo:2022jac}. Such
    constraints will be included in a forthcoming release of {\tt smelli}. As a check, we also mapped onto the EFT bounds on $C_{\ell q}^{(1)}$ from high-$p_T$ Drell-Yan tails~\cite{Allwicher:2022gkm},
    but found the resulting limits to be weaker than those from the bump hunts.}. Since the di-lepton
bump-hunt searches are more sensitive than those of di-jets, we may expect the
FCC-hh to be sensitive to 
all the channels.
For very massive $Z^\prime$ bosons with $M_{Z^\prime} \gg 6$ TeV, one
  could utilise Drell--Yan scattering 
data from the LHC, whose bounds upon SMEFT parameters can be competitive with
those from bump-hunts.

\begin{figure}[!b]
  \centering
\parbox{0.54\textwidth}{\includegraphics[width=0.54\textwidth]{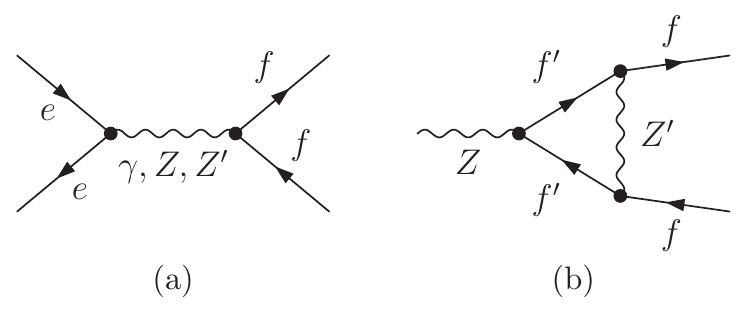}}
\hfill
\parbox{0.4\textwidth}{\vspace{0.3cm}\caption{\label{fig:zpole} Representative
    Feynman diagrams contributing to the $Z$ pole forward-backward asymmetry
    at $e^+ e^-$ colliders (a) and helicity asymmetries (b) at a comparable order in the $M_{Z^\prime}^{-2}$ expansion.}}
\end{figure}

\subsection{$Z^\prime$ at FCC-ee: (Non-)Oblique precision tests}
\label{sec:fcc-ee}
A relatively likely follow-up to the LHC programme in the light of
current ECFA discussions (see
also~\cite{EuropeanStrategyGroup:2020pow}) and potentially imminent
CEPC~\cite{CEPCStudyGroup:2023quu} decisions is an electron-positron
collider, operating at the heavier SM particle mass thresholds, to
produce abundant statistics for high precision $Z$ and Higgs
studies. 
The $Z^\prime$ model discussed in this work leads to modified corrections to
EWPOs in comparison with the SM in two 
categories: oblique and non-oblique corrections, which we shall now discuss in turn.

Firstly, the kinetic mixing of the model implies characteristic
oblique corrections~\cite{Holdom:1985ag} (see
also~\cite{Weihs:2011wp,Choi:2013qra}). These flavour-universal modifications
highlight BSM-induced changes
to SM
relations between the EWPOs
for a chosen
electroweak input parameter set. The kinetic mixing modifies, e.g., the
relation of the $Z$ and $W$ masses for the observed values of $M_Z$, the Fermi
constant $G_F$ and the fine-structure constant $\alpha_{\text{EM}}$. This also
becomes apparent from the matching of the model to the dimension-six Warsaw
basis~\cite{Grzadkowski:2010es} as given in Ref.~\cite{Allanach:2024ozu} with
contributions to $O_{\Phi D}$ linked to the $\hat{T}$
parameter~\cite{Altarelli:1990zd,Peskin:1990zt,Grinstein:1991cd,Altarelli:1991fk}
(also feeding into the forward-backward
asymmetry~\cite{Peskin:1991sw,Burgess:1993vc}). Table~\ref{tab:EWobs} shows the largest contributions to representative observables, based on \emph{current} measurements. The dominant tree-level oblique effect arises from the $\hat{T}$-parameter~\cite{Banerjee:2023qbg,Murphy:2020rsh} (since the model does not match onto the $\hat{S}$-parameter at tree-level), given by
\begin{equation}
\hat{T} = -\frac{v^2}{2 \alpha_{\text{EM}}} C_{\Phi D}.
\end{equation}
To assess future sensitivity, we take projected FCC-ee bounds on $\hat{T}$~\cite{Blondel:2021ema} and map them onto the $Z'$ parameter space using the matching conditions of Ref.~\cite{Allanach:2024ozu}.
\begin{figure}[!b]
  \centering
  \includegraphics[width = 0.6\linewidth]{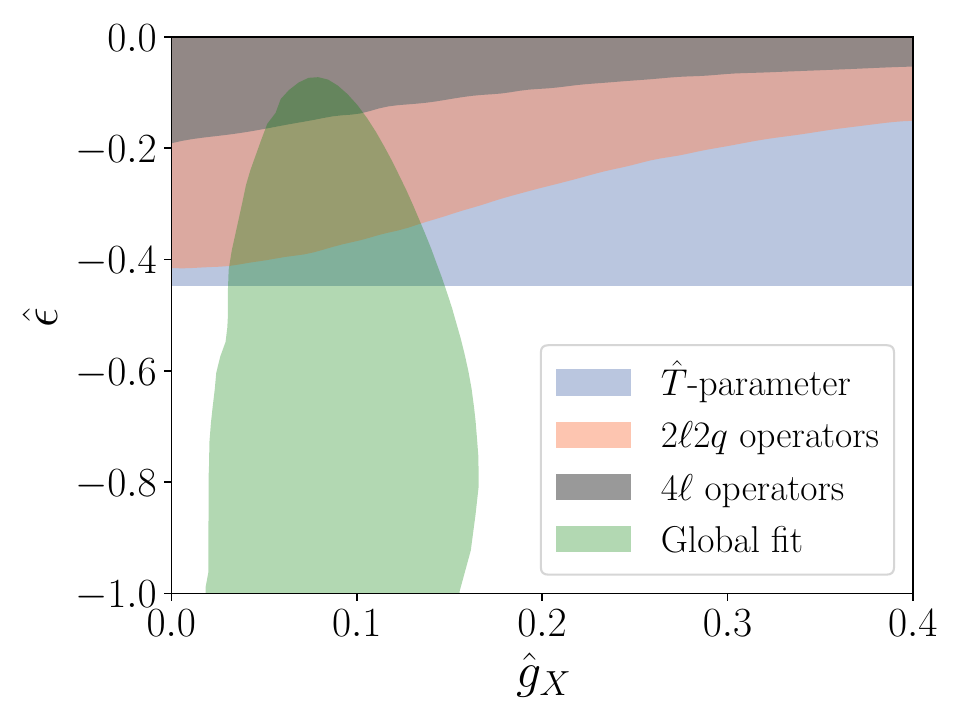}
  \caption{Projected FCC-$ee$ constraints on the model parameter space from the oblique $T$-parameter, and four fermion operators corresponding to $Z^\prime$ mediated $e^+ e^- \to q \bar{q}$ ($2 \ell 2 q$ operators), and $e^+ e^- \to \ell^+ \ell^-$ ($4 \ell$ operators) processes. The shaded regions (excluding the global fit region) represent the FCC-ee sensitive regions at $95 \%$ confidence level. \label{fig:fccee}}
\end{figure}
\begin{table}[!t]
  \begin{center}
    \begin{tabular}{|c|ccc|} \hline
      Observable & experiment & $B_3-L_2$ & SM-pull \\ \hline
      $M_W$/GeV & $80.3795\pm0.01210$ & 80.3646 & 1.60 \\
      $A_{LR}^{ee}$ & $0.15132\pm0.00193$ & 0.14916 & 1.53 \\
      \hline
  \end{tabular}
  \end{center}
  \caption{EWPOs as calculated by {\tt flavio2.3.3} and {\tt smelli2.3.2}~\cite{Aebischer:2018iyb}
    whose SM-pull is larger than unity for the
    best-fit $B_3-L_2$ point and $M_X=3$ TeV, where SM-pull=$|$SM
    prediction $-$ ($B_3-L_2$ prediction)$|$/uncertainty. \label{tab:EWobs}}
  \end{table}
The $Z'$ also generates tree-level contributions to four-fermion interactions, in particular 2-lepton-2-quark ($2 \ell 2 q$) and 4-lepton ($4 \ell$) operators. These contributions are enhanced by the sizeable coupling to electrons, affecting processes such as $e^+ e^- \to \ell^+ \ell^-$ and $e^+ e^- \to b \bar b$. We project FCC-ee sensitivities for the corresponding Wilson coefficients~\cite{Greljo:2024ytg}, including both $Z$-pole and above-$Z$-pole runs, onto the model parameter space using the matching conditions of Ref.~\cite{Allanach:2024ozu}. The combination of high-statistics $Z$-pole data, off-$Z$-pole measurements, and improved $b$-jet tagging yields strong constraints, comparable to the collider bounds shown in Fig.~\ref{fig:collider}. The resulting parameter space, shown in Fig.~\ref{fig:fccee}, indicates that FCC-ee would probe a region similar in reach to the HL-LHC, illustrating the complementarity of precision and direct searches.

Secondly, the tree-level exchange of the new heavy boson will induce a
non-oblique deviation of the $Z$ boson lineshape in exclusive $e^+e^-$
collisions by probing the off-shell $Z^\prime$ contribution and its specific
couplings to, e.g., $b$ quarks and muons. The dominant effect
is driven by the $Z^\prime$ exchange interfering with the SM amplitude
(e.g. Fig.~\ref{fig:zpole}(a)). The relevant effects are $\sim
(\Gamma_Z/M_{Z^\prime})^2 \ll 1$ suppressed ($\Gamma_Z\simeq 2.4~\text{GeV}$ is the
SM width of the $Z$ boson) and therefore require very high precision to obtain
competitive bounds, highlighting the FCC-ee as the relevant environment to
look for such effects.
The expected precision of FCC-ee measurements also motivates the consideration
of loop-suppressed topologies. For measurements on the $Z$ pole, the most
relevant ones are the vertex corrections depicted in
Fig.~\ref{fig:zpole}(b). As they are directly sensitive to the initial and
final state lepton and quark flavours, they provide a further non-oblique
avenue to constrain the malaphoric $B_3-L_2$ model.\footnote{Working in the on-shell scheme, $Z-Z^\prime$ mixing is absent, and additional box topologies are suppressed with respect to the $Z$ boson Breit-Wigner enhancement. In the following, we will not further comment on these effects but interest ourselves in the relative importance of universal and non-universal corrections.} The new ultraviolet singularities sourced by the $Z^\prime$ vertex correction cancel against the fermion wave function corrections in the on-shell scheme, rendering the virtual corrections shown in Fig.~\ref{fig:zpole}(b) manifestly ultraviolet finite for any value of the off-shell $Z$ boson current. Measurable quantities can be derived from the latter by considering the helicity asymmetry ratios of the $Z\to f\bar f$ pole pseudo-observable,
%
\begin{equation}
\label{eq:helicityasym}
A_{\text{LR}}^f = {\Gamma(Z\to f_L\bar f_R) - (R\leftrightarrow L) \over \Gamma(Z\to f_L\bar f_R) + (R\leftrightarrow L)} 
= { \left({1\over 2} - |Q_f| s_w^2\right)^2 - Q^2_f s_w^4 \over \left({1\over 2} - |Q_f|s_w^2\right)^2 + Q^2_f s_w^4 } +
{\cal{O}}\left({m_f^2\over M_Z^2}\right),
\end{equation}
(at leading electroweak order) where $m_f$ is the fermion mass and $Q_f$ is
the electric charge of fermionic field $f$.
It is known that these observables, which are connected to the $e^+e^-\to Z \to f \bar f$ $Z$-pole forward-backward asymmetry via~\cite{Peskin:1995ev}
\begin{equation}
A^f_{\text{FB}} = {3\over 4} A^{e}_{\text{LR}} \, A^f_{\text{LR}},
\label{eq:fb}
\end{equation}
are susceptible to radiative corrections~\cite{Bardin:1986fi,Bardin:1999yd,Baak:2014ora}.
We consider the $Z^\prime$-induced corrections to these observables for different relevant final states $f$ in the following.
%
\begin{table}[!t]
  \centering
  \begin{tabular}{ |c|ccccc|}
      \hline
     & $A_{\text{LR}}^b$ & $A_{\text{LR}}^c$ & $A_{\text{LR}}^\tau$ & $A_{\text{LR}}^\mu$ & $A_{\text{LR}}^e$ \\ 
      \hline
      SM Leading Order & 0.943 & 0.941 & 0.216 & 0.216 & 0.216 \\ 
      SM NLO electroweak & -0.007 & -0.031 & -0.075 & -0.075 & -0.075 \\ 
      \hline
  \end{tabular}
  \caption{Predicted helicity asymmetries in the SM at one-loop electroweak precision. See the main body of text for details.}
  \label{tab:smhelasym}
\end{table}
\begin{figure}[!b]
  \centering
  \includegraphics[width = 0.99\linewidth]{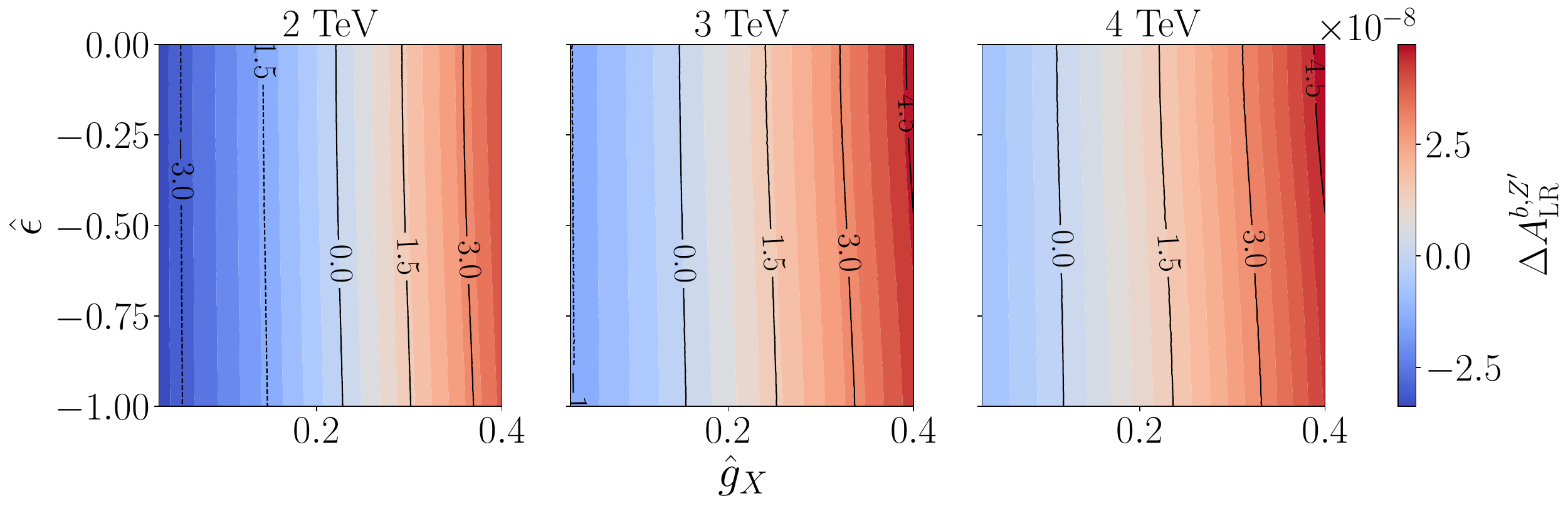}
  \caption{Helicity asymmetry $\Delta A^{b,Z^\prime}_\text{LR}$ modifications induced by the one-loop $Z^\prime$ contribution such as that depicted in Fig.~\ref{fig:zpole}b. Note the tiny scaling factor above the colour bar
    on the far right.\label{fig:ab}}
\end{figure}

\begin{figure}[!t]
  \centering
  \includegraphics[width = 0.99\linewidth]{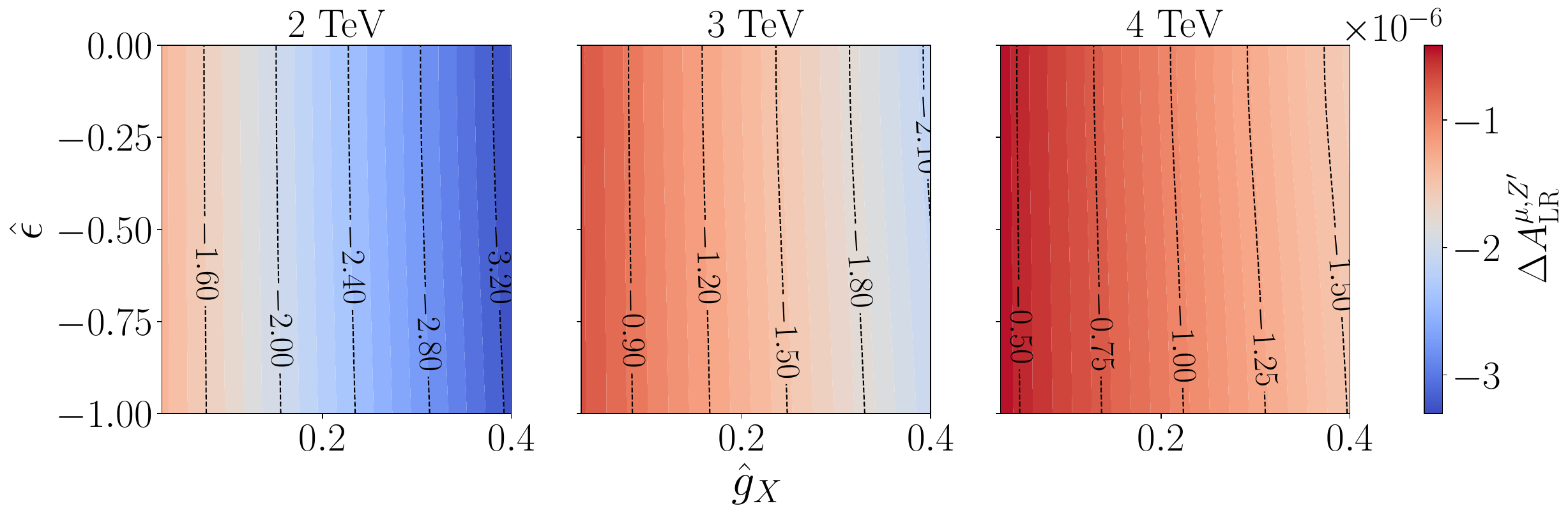}
  \caption{Muon helicity asymmetry $\Delta A^{\mu,Z^\prime}_{\text{LR}}$
    induced by the one-loop $Z^\prime$ contribution such as that depicted in Fig.~\ref{fig:zpole}b. Note the scaling factor above the colour bar
    on the far right.\label{fig:amu}}
\end{figure}

The $Z'$-related corrections should also be compared to the expected size of
the SM weak corrections to qualitatively understand their relevance. Similar
to the computation of the $Z^\prime$ insertion, we have performed a
calculation of $A_{\text{LR}}^f$ in the SM\footnote{From now on, we use input parameters
$\alpha_{\text{EM}} \simeq 1/132.50$, $M_W \simeq 80.42~\text{GeV}$, and $M_Z \simeq 91.18~\text{GeV}$.}. The results\footnote{The soft photon contribution is regularised with an effective photon mass according to Ref.~\cite{tHooft:1978jhc} that cancels against the real emission contribution through the Bloch-Nordsiek~\cite{Bloch:1937pw} or Kinoshita-Lee-Nauenberg theorem~\cite{Kinoshita:1962ur,Lee:1964is}. As the photon emission is non-chiral, its contributions cancel in Eq.~\eqref{eq:helicityasym} except for a finite remainder (see also \cite{Catani:1996vz}).} are shown in Table~\ref{tab:smhelasym}. The comparison clearly indicates that the radiative vertex effects induced by the $Z^\prime$ are small in the parameter region that the HL-LHC explores. 

In the limit of $m_b,M_Z\ll M_{Z^\prime}$ we
find that in e.g.\ the $Z\to b\bar b$ decay the
${\cal{O}}(\epsilon)$ term (for demonstration purposes) is
\begin{equation}
\label{eq:helfra}
\Delta A^{b,Z^\prime}_{\text{LR}}\simeq 
-{\epsilon \,g_X^2  \over 9\pi^2}  {s_w^5(3-2s_w^2)^2 \over (9 - 12 s_w^2 + 8s_w^4)^2}  {M_Z^2\over  M_{Z^\prime}^2}  \left( 6\log\left[m_b^2\over M_{Z^\prime}^2 \right] -1\right) 
\end{equation}
at one-loop order.
This turns out to be small due to the loop suppression factor and the $M_Z^2 /
M_{Z^\prime}^2 \ll 1$ factor. Full one-loop results are presented in Figs.~\ref{fig:ab} and~\ref{fig:amu} for the bottom quark and the muon, respectively.

\begin{figure}[!t]
  \centering
  \includegraphics[width = 0.99\linewidth]{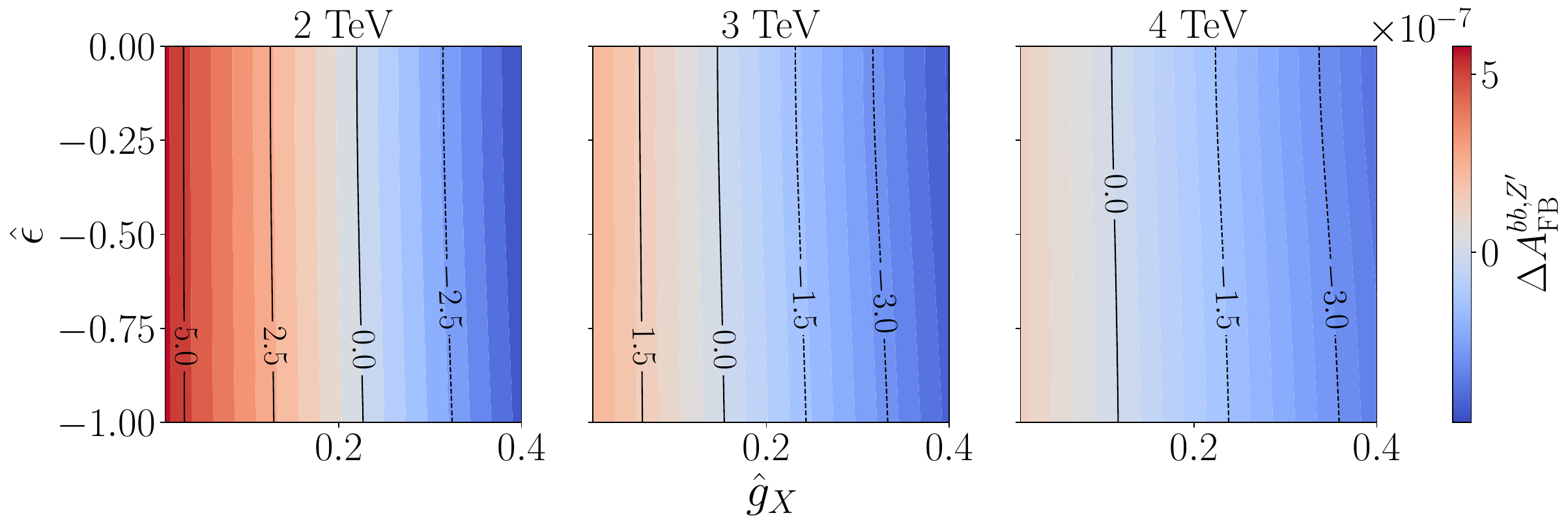}
  \caption{Forward-backward asymmetry $\Delta A_{\text{FB}}^{bb,Z^\prime}$
    modifications induced by the malaphoric $B_3-L_2$ $Z^\prime$  interfering with the $Z$ lineshape.
    Note the scaling factor above the colour bar
    on the right.\label{fig:afbbbtree}}
\end{figure}

In contrast, the interference of the
the tree-level $Z^\prime$ with the SM amplitude, the forward-backward
asymmetry on the $Z$ resonance
(at $M_{Z^\prime} = 2~\text{TeV}$ for example) is
$
  \Delta A^{bb,Z^\prime}_{\text{FB}}   \simeq 3.8\times 10^{-7}
$
for the $b \bar b$ mode for the best fit
point. 

\begin{figure}[!t]
  \centering
  \includegraphics[width = 0.99\linewidth]{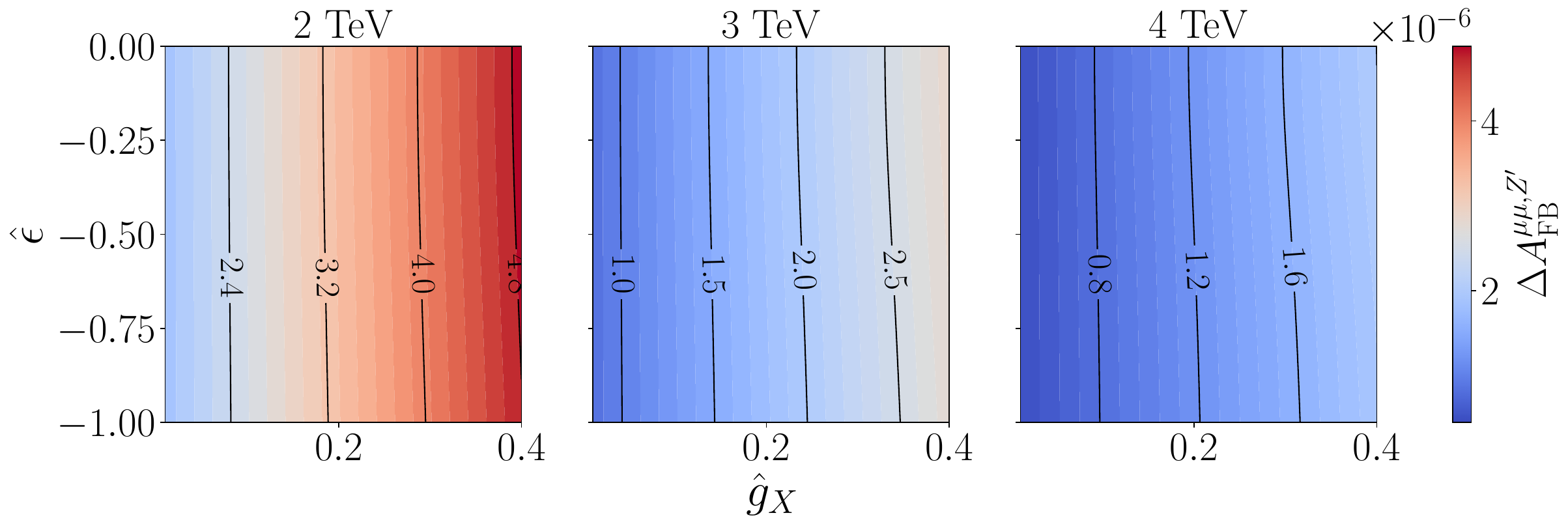}
  \caption{Forward-backward asymmetry $\Delta A_{\text{FB}}^{\mu\mu,Z^\prime}$
    modifications induced by the malaphoric $B_3-L_2$ $Z^\prime$ tree-level
    exchange interfering with the $Z$ lineshape. Note the tiny scaling factor above the colour bar
    on the right.\label{fig:afbmumutree}}
\end{figure}

A scan over the
parameter domain of interest is presented in Figs.~\ref{fig:afbbbtree} and
\ref{fig:afbmumutree}, for the $b$ quark and the muon
for three benchmark points with $M_{Z^\prime} = 2,3,4~\text{TeV}$. 
The flavour-specific contributions discussed above
also compare unfavourably to the projected TeraZ
sensitivity~\cite{Blondel:2019yqr,Blondel:2019jmp,Blondel:2021ema} (e.g., the
estimated uncertainties on $A^{b}_{\text{LR}}$ and $A^{\mu}_{\text{LR}}$ are of
the order $\mathcal{O}(10^{-5})$). 
While the FCC-ee would be able to constrain the model through mixing
effect related EWPOs (see again Tab.~\ref{tab:EWobs}), more direct,
model-specific modifications from loop correction modifications would be
unlikely to provide any additional sensitivity. An FCC-hh, on the other hand,
as shown before, would be sensitive to the whole across the malaphoric
$B_3-L_2$ model's flavour-preferred parameter space. 
\section{Conclusions}
\label{sec:conc}
Flavour measurements inform the search for physics BSM, as do direct exclusion limits from direct new particle
searches at the highest attainable collider energies.
Current measurements in $B-$meson decays that involve the
$b\rightarrow s$ transition
show a significant tension with state-of-the-art SM
expectations~\cite{Gubernari:2023puw,Parrott:2022zte}. We take this as
motivation to consider a particular scenario (the malaphoric $B_3-L_2$ model) that dynamically explains neutral
current anomalies in the $B$ sector to contextualise the anomalies' discovery implications at present and future colliders.

The HL-LHC has the ability to increase sensitivity to the model somewhat
beyond current limits. The oblique corrections have BSM changes to them that would be detectable at the FCC-ee~\cite{Allanach:2024ozu}. FCC-ee non-oblique precision tests, chiefly those informed by characteristic vertex corrections, would unlikely
probe the flavour-preferred region of parameter space beyond this.
We have further shown that the FCC-hh would be sensitive to all of the
flavour-preferred parameter space, even in the less sensitive di-jets
mode.
In the case that the HL-LHC makes no BSM discovery,
at the FCC-hh one could check the correlated signal strengths of equal mass di-electron, di-muon and di-jet bumps in order to further check and constrain
the model.   

\bigskip
\noindent{\bf{Acknowledgements}} --- This work has been partially supported by the Science Technology and Facilities Council consolidated grants
ST/T000694/1 and ST/X000664/1. B.C.A.\ thanks the Cambridge Pheno Working Group for
helpful discussions and the Glasgow Particle Physics
Theory group for hospitality enjoyed while part of this work was carried out. W.N.\ thanks Mario Fernandez Navarro for fruitful discussions.
C.E.\ acknowledges partial support by the Leverhulme Trust Research Project RPG-2021-031 and Research Fellowship RF-2024-300$\backslash$9. C.E.\ is further supported by the Institute for Particle Physics Phenomenology Associateship Scheme.
W.N.\ is funded by a University of Glasgow College of Science and Engineering Fellowship.

\appendix

\section{Definition of matrix $C$ \label{app:c}}
The matrix $C$ defined in Sec.~\ref{sec:model} is given~\cite{Allanach:2024nsa} by the mixing from the original kinetically mixed basis of the
electrically neutral gauge fields to their mass basis:
\begin{equation}
  C = O_{\hat w}^T P_\chi O_{\hat w} O_z, 
\end{equation}
where (in the approximation $M_Z / M_{Z^\prime} \ll 1$)
\begin{eqnarray}
  O_{\hat w} &=& \begin{pmatrix}
    \cos \theta_w & -\sin \theta_w & 0 \\
    \sin \theta_w & \cos \theta_w  & 0 \\
    0        &  0        & 1 \\
  \end{pmatrix},
\qquad
  P_\chi= \begin{pmatrix}
    1 & 0 & - \frac{\epsilon}{\sqrt{1-\epsilon^2}} \\
    0 & 1 & 0 \\
    0 & 0 & \frac{1}{\sqrt{1-\epsilon^2}} \\
  \end{pmatrix}, \\
  O_z &=& \begin{pmatrix}
    1 & 0 & 0 \\
    0 & \cos \theta_z & -\sin \theta_z \\
    0 & \sin \theta_z & \cos \theta_z \\
  \end{pmatrix}, \label{defs}
\end{eqnarray}
and
\begin{equation}
  \tan 2 \theta_z = \frac{-2 M_Z^2 s_w \epsilon \sqrt{1- \epsilon^2}}
  {M_X^2 + M_Z^2 [\epsilon^2 (1+s_w^2)- 1]}.
\end{equation}

\bibliographystyle{JHEP} 
\bibliography{references.bib}

\end{document}